\documentstyle[12pt]{article}
\newcommand{\be}{\begin{equation}}
\newcommand{\ee}{\end{equation}}
\newcommand{\bea}{\begin{eqnarray}}
\newcommand{\eea}{\end{eqnarray}}
\newcommand{\ba}{\begin{array}}
\newcommand{\ea}{\end{array}}

\def\bbox{{\,\lower0.9pt\vbox{\hrule \hbox{\vrule height 0.2 cm
\hskip 0.2 cm \vrule height 0.2 cm}\hrule}\,}}
\newcommand{\dsl}{\pa \kern-0.5em /}

\input epsf.tex

\begin{document}


\begin{titlepage}
\vfill
\begin{flushright}
DAMTP-2003-79\\
hep-th/0308149\\
\end{flushright}

\vfill

\begin{center}
\baselineskip=16pt
{\Large\bf COSMIC ACCELERATION AND M-THEORY\footnote{To appear in the
proceedings of ICMP2003, Lisbon, Portugal, July 2003}}
\vskip 0.3cm
{\large {\sl }}
\vskip 10.mm
{\bf Paul K. Townsend} \\
\vskip 1cm
{\small
Department of Applied Mathematics and Theoretical Physics, \\
Centre for Mathematical Sciences,  University of Cambridge, \\
Wilberforce Road, Cambridge CB3 0WA, UK
}
\end{center}
\vfill
\par
\begin{center}
{\bf ABSTRACT}
\end{center}
\begin{quote}
The status of accelerating four-dimensional universes obtained by
time-dependent compactifications of 10 or 11 dimensional supergravity 
is reviewed, as is the `no-go' theorem that they evade. All flat
cosmologies for a simple exponential  potential are
found explicitly. It is noted that transient acceleration is
generic, and unavoidable for `flux' compactifications. Included is 
an eternally accelerating flat cosmology without a future
event horizon.

\vfill

\end{quote}
\end{titlepage}
\setcounter{equation}{0}

The current consensus that the Universe is undergoing accelerated
expansion presents a challenge to the other current consensus that cosmology
should be derivable from String/M-theory, which has 10 or 11-dimensional
supergravity as its low-energy effective field theory. There are two aspects
to
this challenge. One arises from the fact that there is no known formulation
of
String/M-theory in a spacetime with a future cosmological event horizon
\cite{banks,FS}, whereas this is a typical feature of universes undergoing
late
time eternal acceleration. Of course, we don't know whether the
accelerated expansion of the Universe will
continue forever, so there is no real conflict with observations
here. Still, it remains to find compactifications of String/M-theory for
which
the effective 4-dimensional theory admits  a homogeneous and isotropic
(FLRW)
spacetime undergoing accelerated expansion, and the difficulty of finding
such
compactifications is the other aspect of the challenge posed by cosmic
acceleration. Note that {\sl two} periods of acceleration are required to
explain both inflation in the early universe and acceleration in the current
cosmological epoch.

To see what cosmic acceleration entails, consider, a 4-dimensional
FLRW spacetime in standard coordinates. The metric is
\be
ds^2= -dt^2 + S^2(t) \left[ \left(1-kr^2\right)^{-1} dr^2 + r^2
d\Omega_2^2\right],
\ee
where $S(t)$ is the scale factor and $k=-1,0,1$ depending on whether the
universe
is open, flat or closed, and $d\Omega_2^2$ is the $SO(3)$ invariant metric
on
the unit 2-sphere. A computation of the Ricci tensor shows that
\be
R_{00} = - \partial_t^2 S.
\ee
It follows that an accelerating universe requires $R_{00}<0$. However, the
Einstein field equations imply that
\be
R_{00}= \left(T_{00} + g^{ij}T_{ij}\right)
\ee
and the Strong Energy Condition (SEC) on the matter stress tensor requires
the
right hand side to be non-negative. Thus, accelerated expansion is possible
in a universe governed by Einstein's equations only if the matter in it
violates
the SEC.

{}From a purely 4-dimensional perspective the fundamental condition on the
matter
stress tensor is the Dominant Energy Condition (DEC), which requires $p\ge
-\rho$
for a perfect fluid of pressure $p$ and energy density $\rho$. In contrast,
the
SEC requires only that $p\ge -{1\over3}\rho$, and is typically violated in
theories with a positive scalar potential $V$. For example, given a single
scalar
field $\phi$ and the Lagrangian density
\be\label{4dlag}
{\cal L} = \sqrt{-\det g} \left[{1\over4} R - {1\over2}
\left(\partial\phi\right)^2 -V\right],
\ee
the Einstein equations imply that
\be
R_{00}= 2[\left(\partial_t\phi\right)^2 -V].
\ee
If $V>0$ then $R_{00}$ will be negative whenever $\partial_t\phi=0$,
implying 
an accelerating universe. There will always be {\sl some} cosmological
solutions
for which $\partial_t\phi$ passes through zero. In fact, as will become
clear in
due course, such solutions are the rule rather than the exception! Thus, all
one
needs to get an accelerating universe from String/M-theory is a
compactification
for which the effective 4-dimensional theory has a positive scalar potential
V,
or at least one that is positive in some region of the space of scalar
fields. 

Although the SEC is in no way fundamental, in the sense that its violation
would not
imply a violation of fundamental physical principles, it {\sl is} satisfied
by the
stress tensor of the D=10 and D=11 supergravities that serve as the low
energy
effective theories of String/M-theory\footnote{The `massive' IIA
supergravity is an
exception but, for a reason to be explained later, this exception makes
no difference.}. This fact has consequences for the potential
$V$ in the effective 4-dimensional theories that result from String/M-theory
compactifications, as first pointed out in a 1985 article of Gibbons
\cite{GWG}, and more recently by Maldacena and Nu\~nez \cite{MN}. I will now
summarize
this `no-go theorem'. Consider a D-dimensional spacetime with metric
\be
ds^2_D = \Omega^2(y) ds^2_4(x) + ds^2_n(y)\qquad (n=D-4)
\ee
where $ds^2_4$ is the metric of some 4-dimensional spacetime (with
coordinates
$x$), and 
$ds^2_n$ is the metric of some compact non-singular $n$-manifold ${\cal M}$
(with coordinates $y$). The non-vanishing function $\Omega(y)$ is a `warp
factor'. 
A calculation shows that
\be
R_{00}^{(D)}(x,y) = R_{00}(x) -{1\over4} \Omega^{-2}(y) \nabla^2_y
\Omega^4(y).
\ee
Multiplying by $\Omega^2$ and integrating over ${\cal M}$ we deduce that
\be
\left[\int_{\cal M} \Omega^2\right] \, R_{00} =
\int_{\cal M} \Omega^2 R^{(D)}_{00}\, ,
\ee
and hence that $R_{00}\ge0$ if $R_{00}^{(D)}\ge0$.

This result might appear to rule out the possibility of an accelerating
universe arising
from compactification of 10 or 11-dimensional supergravity. In view of our
earlier
remark that acceleration is always possible when the 4-dimensional scalar
potential $V$
is positive, this would be equivalent to the statement that the potential
arising from
such compactifications is never positive, a statement that is false, and was
known to be false well before the no-go theorem was formulated. In 
fact\footnote{This point was made in \cite{TW}, in a note added to the
published version; it has also been made, independently, in 
\cite{dA}.}, all that can be
inferred about the potential $V$ is that {\sl it has no stationary points
with $V>0$}. 
To see that a positive potential without a stationary point is not excluded,
it
suffices to note that the field equations imply, under the given
circumstances, that
at least one scalar field is time-dependent. This scalar field could be one
that
arises from the mode expansion on the compact manifold ${\cal M}$, in which
case the
metric on
${\cal M}$ will be time-dependent. But the theorem assumed time-independent
${\cal M}$, and is therefore not applicable. One can see from this that the
no-go theorem
is actually a very {\sl weak} constraint on the positive potentials that
might, in
principle, arise from compactification of String/M-theory! There are many
potentials
that it would allow but which, nevertheless, do not seem to be obtainable.
For example,
for many compactifications there is a consistent truncation to a single
scalar field
$\phi$ with a potential of the form
\be\label{exppot}
V= \Lambda e^{-2a\phi} \qquad (\Lambda>0)
\ee
for (dilaton coupling) constant $a$ (which we may assume to be positive).
Any value of
$a$ would be permitted by the no-go theorem but only $a>1$ arises in
practice. This might be expected on the grounds that $a<1$ allows an
eternally accelerating cosmology with a future event horizon. This suggests
the conjecture that such (Einstein conformal frame) 
spacetimes cannot arise from classical compactification of
String/M-theory; if true (there is no known counterexample) this would
impose
much stronger constraints on the potential $V$ than the no-go theorem. In
particular, it would exclude $a<1$ (but not $a=1$, as will be shown later).

Let us now turn to the question of how one gets positive potentials from
(classical) compactification of higher-dimensional theories satisfying 
the SEC. These arise in one of two ways:
\begin{itemize}
\item
{\bf Flux compactifications}: in this case a positive potential is generated
by non-zero flux of antisymmetric tensor fields. The prototype is the $T^7$
compactification of 11-dimensional supergravity with non-vanishing 4-form
field strength \cite{ANT}, which yields an exponential potential of the form
(\ref{exppot}) 
with\footnote{This value, which corresponds to $\Delta=4$ in the
notation of \cite{pktcos} was given incorrectly in that paper. The
correct value is the same as the one found from $T^6$ compactification of
the massive IIA supergravity, which is why the existence of this SEC violating
theory does not alter our conclusions.} 
$a=\sqrt{7}$, the scalar field arising from the `breathing mode' of $T^7$. 
The 4-form is  dual to a 7-form proportional to the volume
form of
$T^7$; more generally, some $k$-form field strength will be set equal to a
closed but
not exact $k$-form on the compact space ${\cal M}$.  Flux compactifications
seem only
to yield `steep' exponential potentials with $a\ge\sqrt{3}$.
\item
{\bf Hyperbolic compactifications}: in this case the compact space is a
space of constant negative curvature. The fact that hyperbolic 
compactifications produce a positive potential was observed by Bremer 
et al. \cite{BDLPS}, and they were investigated by
Kehagias and Russo \cite{russo} in the context of String/M-theory. Several
attractive features (for example, the absence of moduli other than 
the volume) were noted and exploited in a cosmological context by
Kaloper et al. \cite{KMST}, and their possible relevance to cosmic
acceleration was noted by Wohlfarth and the author \cite{TW}. One
could consider the prototype to be the compactification of
11-dimensional supergravity on a 7-dimensional compact hyperbolic 
space. In this case one finds a potential for the breathing-mode 
scalar $\phi$ of the form (\ref{exppot}) with $a= 3/\sqrt{7}$. In
general, hyperbolic compactifications seem always lead to `gentle'
exponential potentials with $1<a<\sqrt{3}$.
\end{itemize}
In general, a positive multi-scalar potential can be generated by a
combination of both mechanisms, in which case it takes the form of a 
sum of products of exponentials of canonically normalized scalar
fields. However, the simple case of a single exponential for a 
single scalar field is sufficient for an understanding of the physics 
and here we shall consider only this case. 

The qualitative features of cosmologies derived from (\ref{4dlag}) with a
potential of the form (\ref{exppot}) were analysed in a 1987 paper of
Halliwell \cite{JJH}, although the fact that there is typically a period of
transient acceleration in the $a>1$ cases was not noticed there. In 2002
Cornalba and Costa \cite{CC} noted the existence of a period of
acceleration in a $k=-1$ cosmology arising from a flux 
compactification\footnote{The acceleration was
claimed to occur in the neighbourhood of a resolved cosmological
singularity. However, for reasons that will hopefully become clear
below, the accelerating epoch is necessarily far from the cosmological 
singularity.}. 
More recently, an explicit time-dependent hyperbolic compactification 
of the vacuum Einstein equations was
shown to yield an Einstein-frame $k=0$ universe in which a decelerating
epoch
with
\be\label{early}
S\sim t^{{1\over3}}, \qquad e^\phi \sim t^{-{1\over\sqrt{3}}},
\ee
is followed by a period of transient acceleration \cite{TW}. This solution
was subsequently shown to be the vanishing flux limit of a rather general class
of solutions of Einstein's equations known as S-branes, and the
phenomenon of transient acceleration was found to be a generic feature of
these solutions \cite{sbranes}.  As observed by Emparan and Garriga
\cite{EG}, this is an immediate corollary of the positive potentials 
generated by flux and hyperbolic compactifications.  Consider the
simple case of an exponential potential of the form (\ref{exppot}) with
$a>1$. The initial conditions implied by (\ref{early}) are $\phi \gg 1$ with
$\dot\phi<0$. Any such cosmological solution can be viewed as a ball
rolling, with friction, up the potential. Clearly, it must reach a
maximum at which $\dot\phi=0$ and at this point the expansion of the 
universe is accelerating, for the reason explained
previously. Subsequently, the ball starts to roll back down the
hill; the late-time behaviour will depend on the value of $a$ and 
also on $k$, but in all cases the universe will be decelerating. For 
example, for $a^2<3$ and $k=0$ the late-time behaviour will be given
by the power-law k=0 attractor solution
\be\label{fixed}
S \sim t^{1/a^2} ,\qquad e^\phi \sim t^{1/a}.
\ee
Nearby trajectories with $k\ne0$ will eventually approach a Milne universe
attractor or collapse to a big crunch singularity. Note that the compact
Kaluza-Klein space ${\cal M}$ starts at infinite volume and ends at infinite
volume; the acceleration of the 4-dimensional cosmology is associated
to a `bounce' of the compact space off its minimal volume.

The above explanation of the period of transient acceleration relies only on
the positivity of the potential $V$ and makes no distinction between flux
compactifications and hyperbolic compactifications. However, Halliwell's
analysis \cite{JJH} shows that just as the `critical' value $a=1$ separates
qualitatively
different behaviours of cosmological trajectories in the class of models
under
discussion, so does the `hyper-critical' value $a=\sqrt{3}$, which also
separates
hyperbolic from flux compactifications. To see this, we introduce a new time
parameter $\tau$ such that
\be
d\tau = e^{-a\phi} dt,
\ee
and set $S= e^{\alpha(\tau)}$.
Letting an overdot indicate differentiation with respect to $\tau$, we
find that the $\phi$ equation of motion is
\be
\ddot \phi -a \dot\phi^2 +  3\dot\alpha \dot\phi = 2a,
\ee
while the Friedmann equation is
\be
\dot\alpha^2 - {1\over3}\dot\phi^2 = {2\over3} - {ke^{2a\phi} \over S^2}.
\ee
{}For $k=0$, the above two equations are equivalent to
\be\label{two}
\ddot\phi = 3\dot\alpha \left(a\dot\alpha - \dot\phi\right),
\ee
and
\be\label{one} 
3\dot\alpha^2 - \dot\phi^2 = 2,
\ee
which is a hyperbola separating the the $k=-1$ and $k=+1$ trajectories. The
$\dot\alpha>0$ branch of the hyperbola corresponds to an expanding
universe. We can parametrize this branch by writing
\be\label{alphi} 
\dot\alpha = {1\over \sqrt{6}}\left(\xi + \xi^{-1}\right) ,\qquad 
\dot\phi = {1\over \sqrt{2}} \left(\xi - \xi^{-1}\right), \qquad (\xi>0).
\ee
Equation (\ref{two}) then becomes
\be
\label{ODE}
\dot\xi = {1\over\sqrt{2}}\left[ \left(a+\sqrt{3}\right) +
\left(a-\sqrt{3}\right)\xi^2\right].
\ee

We now see why $a=\sqrt{3}$ is special. For $a<\sqrt{3}$ there is a
fixed point solution
\be
\xi = \xi_0 \equiv \left({\sqrt{3} +a \over \sqrt{3}
-a}\right)^{1\over2},
\ee
which is just the power-law attractor solution (\ref{fixed}). 
The fixed point separates two other $k=0$ trajectories:
\be
(i)\ \xi = \xi_0 \coth \gamma\tau \qquad 
(ii) \ \xi = \xi_0 \tanh \gamma \tau, \qquad 
\gamma = \left({3-a^2\over2}\right)^{1\over2}
\ee
where $\tau>0$. Only case (ii) includes $\xi=1$, and 
hence $\dot\phi=0$; this solution
undergoes a period of acceleration whereas the other does not. 
In either case one can integrate (\ref{alphi}). In case (i) one has
\be
S^{\sqrt{3}} \propto \left(\cosh\gamma\tau\right)^{\lambda_+}
\left(\sinh\gamma\tau\right)^{\lambda_-}, \qquad
e^\phi \propto \left(\cosh\gamma\tau\right)^{-\lambda_+} 
\left(\sinh\gamma\tau\right)^{\lambda_-}, 
\ee
where
\be
\lambda_\pm = {1 \over \sqrt{3} \pm a}.
\ee
There is a big-bang singularity at $\tau=0$, near
which
\be
S \sim t^{1\over 3}, \qquad e^\phi \sim t^{1\over \sqrt{3}}, 
\ee
so the volume of the compact Kaluza-Klein space ${\cal M}$ is
initially zero. Subsequently the solution approaches the attractor
(\ref{fixed}). In case (ii) we have
\be
S^{\sqrt{3}} \propto \left(\cosh\gamma\tau\right)^{\lambda_-}
\left(\sinh\gamma\tau\right)^{\lambda_+}, \qquad
e^\phi \propto \left(\cosh\gamma\tau\right)^{\lambda_-} 
\left(\sinh\gamma\tau\right)^{-\lambda_+}.
\ee
This behaves intially as in (\ref{early}) but then passes through a
period of acceleration before approaching the attractor (\ref{fixed}).

Both the above solutions were found in \cite{TW}, for particular
values of $a$ in the range $1<a<\sqrt{3}$, as solutions of the
$D\ge6$ vacuum Einstein equations with a compact hyperbolic $D-4$
dimensional manifold of time-dependent volume. As solutions of the
4-dimensional effective theory with Lagrangian density (\ref{4dlag}), 
they are actually valid for $0\le a <\sqrt{3}$, in particular for $a=1$.
The power-law attractor solution in this case has $S\sim t$ and hence
zero aceleration, so the case (ii) solution that approaches it
asymptotically must be {\it eternally} 
accelerating\footnote{A similar observation 
was made in \cite{CHNOW} in the $a>1$ case for
trajectories that approach the $k=-1$ Milne attractor.}.
In fact, the late time behaviour is
\be
e^\phi \sim  t + {2\sqrt{3}\over t} + {\cal O}(t^{-2}), \qquad
S \sim t + {4\over \sqrt{3} t} + {\cal O}(t^{-2}).
\ee
from which one sees that $\partial_t^2S>0$. One might suspect from 
this fact that there
would be a future cosmological event horizon, in which case 
$a=1$ would be excluded by the (admittedly conjectural) stronger 
form of the no-go theorem proposed earlier in this article.
However, it has been shown by Boya et al. \cite{boya} that if the
acceleration tends to zero asymptotically, as it does in this case, 
then there is no cosmological event horizon. 

Halliwell's qualitative analysis of all cosmological trajectories can
similarly be made quantitative for $k=0$ when $a\ge3$. Consider first
the $a>\sqrt{3}$ case. The solution of (\ref{ODE}) is
\be
\xi = \left({a+ \sqrt{3}\over a-\sqrt{3}}\right)^{1\over2} \tan \omega
\tau, \qquad \omega = \left({a^2-3\over2}\right)^{1\over2}, 
\ee
where $0<\tau<\pi/2$. The equations (\ref{alphi}) can 
now be integrated to yield
\be
S^{\sqrt{3}}\propto \left(\cos\omega\tau\right)^{\lambda_-}
\left(\sin\omega\tau\right)^{\lambda_+}, \qquad 
e^\phi \propto \left(\cos\omega\tau\right)^{\lambda_-}
\left(\sin\omega\tau\right)^{-\lambda_+}.
\ee
The aymptotic behaviour as $\log t\rightarrow \pm \infty$ is
\be
S\sim t^{1\over3}, \qquad e^\phi \sim t^{\pm 1/\sqrt{3}}.
\ee
In between there is a period of acceleration. For $a=\sqrt{3}$ we have
simply $\xi = \sqrt{6}\tau$ and hence
\be
S^3 \sim \tau^{1\over2} e^{{3\over2}\tau^2}, \qquad e^{\sqrt{3}\phi} \propto
\tau^{-{1\over2}} e^{{3\over2}\tau^2}.
\ee
The late time behaviour now involves logarithmic corrections to the
power law behaviour of the $a>\sqrt{3}$ case. 

Despite the differences between flux compactifications and hyperbolic
compactifications, all the above cases of accelerating $k=0$ universes are
qualitatively similar. There is always a big-bang singularity 
near which the scale factor behaves as in
(\ref{early}). Inspection of the phase portraits in \cite{JJH} shows 
that this is also true for the $k\ne0$ cosmologies. This is not 
surprising because the singularity theorems that guarantee
a cosmological singularity  rely on the SEC which is violated by the
4-dimensional effective theory only in a {\sl later} epoch. 
In addition, the acceleration leads to only a few
e-foldings, insufficient for any application to inflation in the early 
universe; this disappointing conclusion is confirmed by
systematic analyses of the possibilities of hyperbolic-flux
compactifications involving many scalar fields
\cite{CHNOW,Wohl3}. However, the mechanism may be
relevant to acceleration in the current cosmological epoch \cite{GKL}. One
of its attractive features is that any time-dependence of four-dimensional
`constants' due to the time-dependence of the compact space is absent 
precisely during the period of acceleration, i.e. now!

Implicit in everything discussed so far in this article is the assumption
that String/M-theory is adequately described for cosmological purposes by the
classical effective 10 or 11 dimensional supergravity theories. It
seemed worthwhile to fully explore the implications of this
assumption, but it now also seems that it must be discarded because 
although it has been possible to find models with
transient acceleration of possible relevance to the `observed' 
acceleration of the current cosmological epoch, it has not been
possible to find models that allow a sufficient period of early
universe inflation. This is not a disaster; String/M-theory
includes orientifolds that violate (at least locally) the SEC \cite{GKP}, and
branes which can produce brane-instanton quantum corrections to the 
potential $V$. It has been shown \cite{KKLT} that when these effects
are taken into account it is possible for
the potential $V$ to have a local minimum that could lead to
inflation\footnote{Such a potential would allow classical solutions with a
future cosmological event horizon, but since quantum mechanics has now been
invoked one can presumably invoke it again to argue that the universe will
eventually tunnel out of the metastable vacuum.}.
Nevertheless, although string and brane effects might yield potentials $V$
that are quite different from those obtainable from the classical 
compactifications of effective supergravity theories considered here, 
some of the lessons learned from the latter case may prove
valuable. If one re-interprets $a$ in (\ref{two}) and
(\ref{one}) to be shorthand for $V'/2V$ then these equations describe the
$k=0$ cosmologies for {\sl any} positive potential $V$. As long as 
$a\ge \sqrt{3}$, there will be no fixed point on the $k=0$ hyperbola and 
hence a single $k=0$ trajectory, which will necessarily include a
period of acceleration. Thus, for a large class of potentials,
including all those that arise classically from `product' 
flux compactifications, an
accelerating epoch is not only possible, but (assuming a flat
universe) unavoidable! If $a< \sqrt{3}$ at some point then fixed
points will occur, and acceleration can be avoided. But note that
we are now discussing how to avoid acceleration rather than how to achieve
it! When it was noted in \cite{TW} that hyperbolic compactification can yield
both a flat 4-dimensional universe that undergoes a period of 
acceleration {\sl and} a flat 4-dimensional universe that is always 
decelerating, it was the former case that appeared remarkable, but 
actually it is the latter case, an eternally decelerating 
universe, that is exceptional. Of course, the acceleration is transient, so
even if it is generic one could ask why we happen to be around to
observe it. But is it really any more surprising that we live at an 
atypical time in a typical universe than that we
live at a typical time in a atypical universe?

\bigskip
\noindent
{\bf Acknowledgements.} The author thanks Michael Douglas, Gary
Gibbons, Jaume Garriga, Jaume Gomis, Shamit Kachru and Mattias 
Wohlfarth for helpful discussions.

\end{document}